# Optical Wireless Communication Systems, A Survey

Osama Alsulami[1], Ahmed Taha Hussein[1], Mohammed T. Alresheedi[2] and Jaafar M. H. Elmirghani[1]
[1]School of Electronic and Electrical Engineering, University of Leeds, LS2 9JT, United Kingdom
[2]Department of Electrical Engineering, King Saud University, Riyadh, Kingdom of Saudi Arabia
ml15ozma@leeds.ac.uk, asdftaha@yahoo.com, malresheedi@ksu.edu.sa, j.m.h.elmirghani@leeds.ac.uk

*Abstract*—In the past few years, the demand for high data rate services has increased dramatically. The congestion in the radio frequency (RF) spectrum (3 kHz ~ 300 GHz) is expected to limit the growth of future wireless systems unless new parts of the spectrum are opened. Even with the use of advanced engineering, such as signal processing and advanced modulation schemes, it will be very challenging to meet the demands of the users in the next decades using the existing carrier frequencies. On the other hand, there is a potential band of the spectrum available that can provide tens of Gbps to Tbps for users in the near future. Optical wireless communication (OWC) systems are among the promising solutions to the bandwidth limitation problem faced by radio systems. In this paper, we give a tutorial survey of the most significant issues in OWC systems that operate at short ranges such as indoor systems. We consider the challenging issues facing these systems such as (i) link design and system requirements, (ii) transmitter structures, (iii) receiver structures, (iv) challenges and possible techniques to mitigate the impairments in these systems, (v) the main applications and (vi) open research issues. In indoor OWC systems we describe channel modelling, mobility and dispersion mitigation techniques. Infrared communication (IRC) and visible light communication (VLC) are presented as potential implementation approaches for OWC systems and are comprehensively discussed. Moreover, open research issues in OWC systems are discussed.

## I. INTRODUCTION

TRADITIONAL radio communication systems suffer from limited channel capacity and transmission rate due to the limited radio spectrum available, while the data rates requested by the users continue to increase exponentially. Different techniques have been proposed to use the radio spectrum in more efficient ways including advanced modulation, coding, smart antennas and multiple input and multiple output (MIMO) systems [1], [2].

Achieving very high data rates (beyond 10 Gbps and into the Tbps regime) using the bandwidth of radio systems is challenging. According to Cisco, mobile Internet traffic over this half of the decade (2016-2021) is expected to increase by 27 times [3]. Given this expectation of dramatically growing demand for data rates, the quest is already underway for alternative spectrum bands beyond radio waves. The latter are bands currently used and planned for near future systems, such as 5G cellular systems.

Different technology candidates have entered the race to provide ultra-fast wireless communication systems for users, such as optical wireless communication (OWC) systems for indoor usage [4]–[6].

OWC systems are candidates for high data rates in the last mile of the access network. They have been investigated for more than three decades as an alternative solution to support high speed data instead of radio frequency (RF) systems [4]. However, this technology was not deployed until the 1990s when the transmitter and the receiver components became available at low cost [7]. Their rate of adoption is expected to grow now in view of the use of semiconductor sources in lighting applications indoor and in view of the significant growth witnessed in data rates. In OWC systems, a modulated beam of infrared, visible or ultra violet light is transmitted, typically through the atmosphere, by an OW transmitter. OWC systems use light emitting diodes (LEDs) or laser diodes (LDs) to transmit light similar to what is done in fibre optic systems.

Indoor OWC systems include systems such as those standardized by the infrared data association (IrDA), and systems developed to support infrared optical wireless communication (IRC) and visible light communication (VLC) [8].

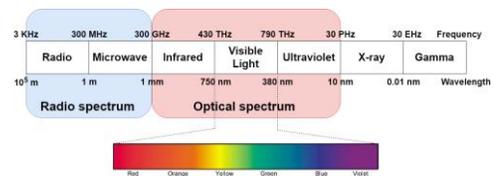

Fig. 1: The Electromagnetic spectrum.

OWC systems have been shown in many studies to be able to transmit video, data and voice, at data rates up to 25 Gbps in indoor systems [9], [10], [11], [12], [13].

OWC systems offer many favourable features when compared to RF systems, such as rapid deployment, low start-up and operational costs and higher bandwidth, similar to fibre optics. The available OW spectral regions offer virtually unlimited bandwidth, such as 1 mm – 750 nm for infrared, 780 nm – 380 nm for visible light and 10 nm – 400 nm for UV as shown in Fig. 1. However, there are certain wavelengths that are suitable for communication in OW spectrum such as the range between 780 – 950 nm which is the optimal range considered in IRC and the ultraviolet-C band, which is the best range for wide field of view (FOV) communication [14].

Since, the OW spectrum is unregulated and is licence free, the cost of the OW systems can be reduced compared to RF systems.

The main features and drawbacks of OWC systems are presented here. The nature of light gives OWC systems immunity against interference caused by adjacent channels in neighbouring rooms, with the possibility of frequency reuse in different parts of the same building, which means increased capacity. OWC systems also offer better security at the physical layer. This is due to the fact that light does not penetrate opaque barriers, which means the potential for eavesdropping is reduced unlike in radio systems, and this reduces the need for data encryption [5], [15]. In addition, the detector (photodiode) has a very large area, typically tens of thousands of wavelengths, and this leads to efficient spatial diversity at the receiver [15]. These combined favourable features make OWC systems suitable replacements for conventional radio systems. However, OWC systems have several drawbacks. In indoor systems, optical wireless access points, in different rooms, have to be connected via a wired backbone as light cannot penetrate walls. Also, because of multipath propagation, reflections and signal spread, optical signals suffer from attenuation and dispersion leading to inter symbol interference (ISI). Furthermore, the signal at the receiver can include shot noise induced by intense ambient light sources (sunlight, incandescent lighting and fluorescent light), and this leads to signal corruption by this background noise [16], [17]. Moreover, the transmitter power is limited by eye and skin safety regulations [18], [19].

OWC systems require a receiver with a photo sensitive detector with large area to collect the maximum optical signal possible, and this leads to an increase in the capacitance of the photodetector, consequently the available receiver bandwidth decreases.

Table 1 compares Radio and OW systems. To place OWC in the communications scene, Fig. 2 summarizes the coverage area and data rates of the available systems in traditional RF and OWC systems.

Table 1: Comparison of radio and OW systems

| Property | Radio System | OW System |
|---|---|---|
| **Bandwidth Regulated** | YES | NO |
| **Security** | Low | High |
| **RF Electromagnetic interference** | YES | NO |
| **Passes through walls** | YES | NO |
| **Technology Cost** | High | Low |
| **Beam Directionality** | Low | Medium |
| **Available bandwidth** | Low | Very high |
| **Transmitted Power** | Restricted (Interference) | Restricted (Eye safety and interference) |
| **Noise Sources** | Other Users and Systems | Sun light and Ambient Light |
| **Power consumption** | Medium | Relatively low |
| **Multipath Fading** | YES | NO |

OWC systems can provide high data rates beyond 10 Gbps [9], [11], [12], [13], however, there are many challenges to implement OW systems. For instance, fog, rain and dust reduce data rates and coverage area of free space optics (FSO) outdoor systems, while multipath propagation, receiver noise and interference tend to limit the capacity of indoor OWC systems.

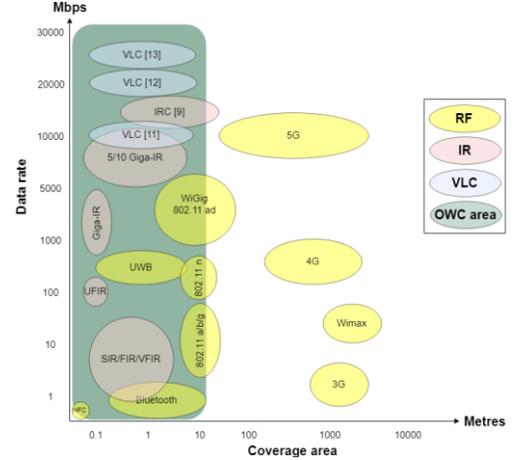

Fig. 2: Data rates and coverage area of different wireless technologies [9], [11], [12], [13], [20].

The rest of the paper is organized as follows: Section II presents a discussion of link design, modulation techniques and the channel characteristics of OWC systems. Section III discusses the available transmitter architectures. Section IV summarizes the OW receiver structures and hence the design of OWC systems. In Section V, the main challenges faced in OWC systems are described. Due to the importance of VLC systems, we provide a comprehensive review of VLC systems in a separate Section VI. Section VII summarizes the main applications of OWC (IRC and VLC) systems. Section VIII gives an overview of the open research issues in IRC and VLC systems. Finally, we conclude the paper in Section IX.

## II. LINK DESIGN, CHANNEL CHARACTERISTICS AND MODULATION TECHNIQUES IN OW SYSTEMS

In this section, we discuss various link designs that can be employed in OWC systems. A brief description of the modulation techniques and channel characteristics of indoor OWC system is also provided.

### A. Link design

There are two criteria to classify OWC links: (a) the level of directionality of the receiver and the transmitter and (b) if there is a direct path between the receiver and the transmitter. These classifications are based on the radiation pattern of the transmitter and field of view (FOV) of the receiver. Link classification schemes are illustrated in Fig. 3. The two most common link configurations for indoor systems are line of sight (LOS) and non-LOS (NLOS) [5], [9], [21], [22]. LOS links provide a direct path between the transmitter and the receiver which minimizes multipath dispersion and enhances the power efficiency of the communication system. However, LOS links suffer from shadowing. On the other hand, NLOS links rely on reflections from the walls, ceiling and other objects. They offer robust links and protection against shadowing but are severely affected by multipath dispersion, which results in ISI and pulse spread [22].

Both LOS and NLOS schemes can be classified into directed, hybrid and non-directed links [22]. The NLOS non-directed scenario is considered to be the most desirable method for indoor systems from the point of view of mobility, however it suffers dispersion which can limit the achievable data rate and is also limited by the low power collected. A very popular diffuse system is the conventional diffuse system (CDS) [22]. CDS employs a wide FOV receiver and a transmitter with wide radiation pattern. The transmitter and receiver point towards the ceiling and the received signal reaches the receiver through multiple reflections. The CDS impairments can be reduced by using specific receivers, such as the triangular pyramidal fly-eye diversity receiver which uses an optimum value of the FOV to mitigate ISI in indoor OWC systems [21]. However, non-directed links do not have good power efficiency. Thus, employing different degrees of directionality at the transmitter and receiver (hybrid link) can give good performance [22].

LOS communication using directed links makes use of a narrow radiation pattern between the transmitter and the receiver (FOV), see Fig. 3 [22]. This scenario reduces the impact of ambient noise and minimizes the path loss, while also improving the power efficiency.

Directed links are however not suitable for mobile applications especially for indoor OWC usage. In contrast, non-directed links use a wide transmitter radiation pattern and wide FOV at the receiver. Therefore this configuration can provide mobility for users (see Fig. 3).

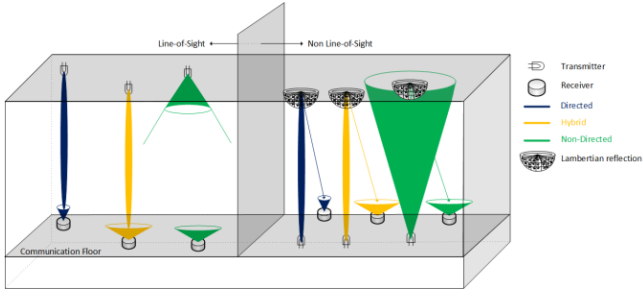

Fig. 3: Optical wireless link classification

B. Channel characteristics and modulation techniques

This section discusses the indoor OW channel characteristics and the popular modulation schemes used.

1. Channel characteristics

Channel characterization in OWC is vital to better understand the system performance. The channel impulse response $h(t)$, which can fully describe the multipath propagation in the OWC system, is given by [23]:

$$y(t, Az, El) = \sum_{k=1}^{K} Rx(t) \otimes h_k(t, Az, El) + R \sum_{k=1}^{K} N_k(t, Az, El) \qquad (2)$$

where $K$ is the total number of reflecting elements in the room, $Az$ and $El$ are the directions of arrival in the azimuth and elevation angles, respectively. By estimating the impulse response, many parameters can be determined, such as the root mean square, delay spread, 3 dB channel bandwidth, spatial power distribution and signal to noise ratio (SNR), [23].

To give examples of the OW channel impulse response, we show the impulse responses of CDS and of line strip multi beam system (LSMS). The latter is a NLOS hybrid system where the transmitter faces up and produces a number of spots on the ceiling [23]. In addition, the impulse response when using different types of receiver configurations in VLC systems are shown. A ray-tracing algorithm simulation package was developed using the same room configuration and parameters used in [21], [23], [24], [25] and the simulation scheme was based on [26]. The room has length × width × height of $8m \times 4m \times 3m$. The devices used for communication are placed on a "communication floor", ie typical desk surfaces that are one meter above the floor. In the IRC system, the transmitter and receiver were located at the center of the room on the communication floor, and the receiver specifications for the IRC example were based on [27]. In the VLC system 8 light units were used as transmitters and each light unit has 9 LEDs where the light units are located in different positions on the ceiling and the receiver is located at the room center on the communication floor. The locations of the light units, room dimensions (same as above) and receiver specifications of the VLC system are based on [24], [25]. Fig. 4A illustrates the impulse responses of the most well-known indoor IRC systems (CDS, LSMS), Fig. 4B shows the impulse response of a VLC system with a single wide field of view (WFOV) receiver, Fig. 4C shows the impulse response of a VLC system with three branch angle diversity receiver (ADR) and Fig. 4D shows the impulse response of the VLC system using a 50 pixel imaging receiver.

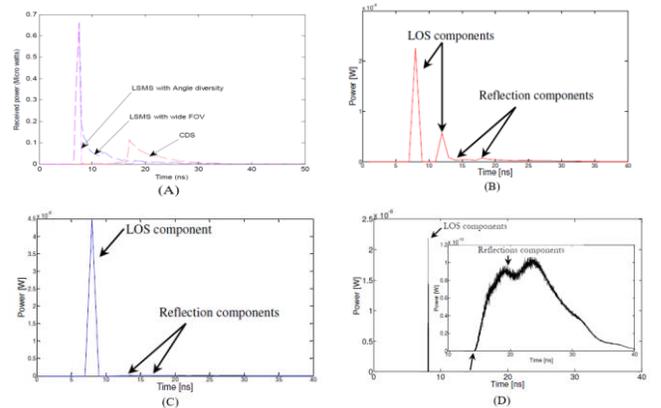

Fig. 4: (A) Impulse responses of CDS and LSMS systems with single element wide FOV receiver ($FOV = 90°$) and LSMS systems with three branches ADR; transmitter and receiver are located at the room center, (B) Impulse responses of VLC system with single wide FOV receiver, (C) Impulse responses of VLC system with three branch ADR, (D) Impulse responses of VLC system with 50 pixels imaging receiver.

Fig. 5 (A) shows that the CDS has the highest delay spread and hence the smallest channel bandwidth compared to the other systems. The LSMS system results in a lower delay spread due to the presence of the diffusing spots which act as secondary transmitters close to the receiver. Also as Fig. 5 (A) shows, reducing the receiver FOV, results in the collection of a smaller range of rays and hence a larger channel bandwidth.

Table 2 provides a comparison of the OW channel bandwidths achievable using different OW systems[24]. The results are shown at $x=2m$ and along the y axis. The minimum channel bandwidth is observed in the case of the CDS with wide FOV receiver, and is 41 MHz in the $8m \times 4m \times 3m$ room using the parameters in [24]. The use of an imaging receiver in conjunction with the CDS increases the minimum channel bandwidth to 160 MHz due to the use of narrow FOV pixels in the imaging receiver which reduce the range of rays collected. The imaging receiver however provides wide coverage through its many pixels which also result in increased collected power. The use of an LSMS transmitter results in multiple diffusing spots cast on the ceiling, which act as secondary emitters close to wherever the receiver is located. This increases the channel bandwidth. The use of power adaptation, where the spot closest to the receiver is allocated more power, reduces the delay spread further, resulting in a minimum channel bandwidth of 7.5 GHz as Table 2 shows.

Table 2: Comparison of the 3dB channel bandwidth of different OW systems [28]

| Configuration | 3 dB Channel Bandwidth (GHz) | | | | | | |
|---|---|---|---|---|---|---|---|
| | Y (m) | | | | | | |
| | 1 | 2 | 3 | 4 | 5 | 6 | 7 |
| CDS with a single non-imaging receiver (FOV=130°) | 0.064 | 0.084 | 0.071 | 0.067 | 0.068 | 0.056 | 0.041 |
| CDS with a single imaging receiver-SB | 0.31 | 0.3 | 0.29 | 0.2 | 0.19 | 0.18 | 0.16 |
| LSMS with a single imaging receiver-SB | 0.83 | 0.75 | 0.62 | 0.46 | 0.34 | 0.22 | 0.19 |
| ALSMS with a single imaging receiver-SB | 7.61 | 7.58 | 7.57 | 7.56 | 7.54 | 7.51 | 7.5 |

Figs. 5 (B) and (C) show the delay spread of VLC systems with wide FOV receiver, angle diversity receiver (ADR) and imaging receiver. Both the ADR and imaging receiver employ multiple receivers with smaller FOV which reduces the delay spread. Full coverage of the room is achieved through the use of the multiple receivers in ADR and the many pixels in the imaging receiver. The smaller FOV of each receiver element in the imaging receiver used, results in reduced delay spread compared to ADR, as Figs 5 (B) and (C) show.

The VLC system with an imaging receiver provides low delay spread as shown in Fig. 5. It also provides good SNR at high data rates compared to the other systems as shown in Fig. 6. The oscillation in delay spread is due to the proximity (or otherwise) from the transmitters (lighting sources in VLC) as the receiver moves along the *y* axis. The oscillations in the SNR are due to similar effects and can be more significant in infrared systems where the room lighting can act as a noise source. The use of optimum receiver element selection or receiver element signal combining techniques in ADR [24] and imaging receivers [24], [25] can result in more uniform SNR in the room where the best receiver element that avoids noise and/or interference for example is selected.

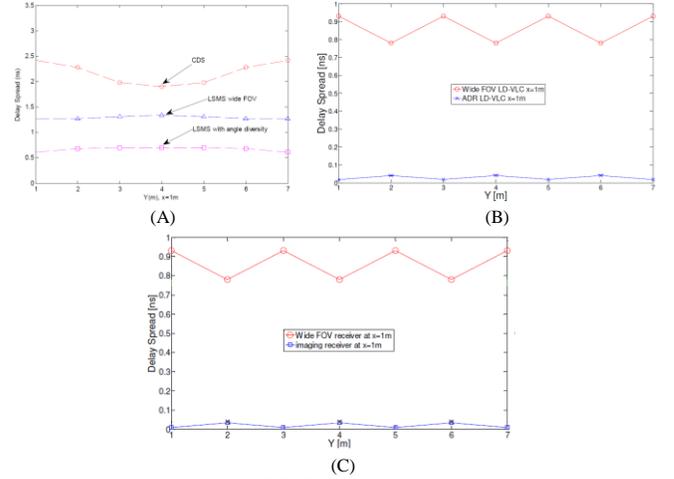

Fig. 5: (A) Delay Spread of CDS and LSMS systems with single element wide FOV receiver (90°) and LSMS system with three branch ADR; transmitter and receiver are located at the room center, (B) Delay Spread of VLC system with single wide FOV receiver and three branch ADR, (C) Delay Spread of VLC system with single wide FOV receiver and 50 pixels imaging receiver.

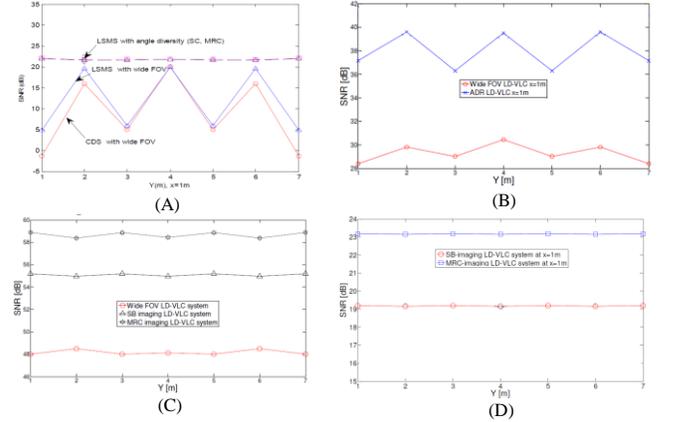

Fig. 6: (A) SNR of CDS and LSMS systems with single element wide FOV receiver (90°) and LSMS system with three branch ADR; transmitter and receiver are located at the room center, (B) SNR of VLC system with single wide FOV receiver and three branch ADR, (C) SNR of VLC system with single wide FOV receiver and 50 pixels imaging receiver at low data rate, 30 Mbps, (D) SNR of 50 pixels imaging receiver at high data rate, 5 Gbps.

2. Modulation techniques

OWC channels are different from traditional RF channels. This has resulted in different methods of modulation being used. Modulation schemes that fit well in RF channels do not necessarily perform well in the optical domain. There are four criteria that guide the choice of a specific modulation technique for OWC systems. Of particular importance as a criterion to be applied, is the average power used (power efficiency) of a given modulation format. This is important in view of eye hazards and power consumption in mobile terminals. The second criterion is the available channel bandwidth and receiver bandwidth requirements. The third factor is the complexity of the modulation format (and power consumption in portable devices). The last set of factors relate to the physical limitations in the transmitter (i.e. LD or LED) which the modulation format may have to take into account. Modulation in OWC systems consists of two steps: in the first step the information is coded as a waveform(s) and then (second step) these waveforms are modulated onto the instantaneous power of the carrier [22].

Intensity modulation (IM) and direct detection (DD) is the preferred transmission technique in OWC systems [9], [22]. IM is achieved by varying the bias current of the LD or the LED. In OWC systems, the transmitted signal must always be positive in intensity. Direct detection is the simplest method that can be used to detect an intensity modulated signal. The photodetector generates a current that is proportional to the incident optical power intensity. A simple description for the IM/DD channel is given as [22]:

$$y(t) = Rx(t) \otimes h(t) + Rn(t) \qquad (1)$$

where $R$ is the photodetector responsivity, $y(t)$ is the instantaneous photo current received, $t$ is the absolute time, $\otimes$ denotes convolution, $h(t)$ is the channel impulse response, $x(t)$ is the instantaneous transmitted power and $n(t)$ is the background noise (BN), which is modelled as white Gaussian noise, and is independent of the received signal.

Three broad types of modulation schemes can be applied in OWC (a) baseband modulation, (b) multicarrier modulation and (c) multicolor modulation [14], [29].

a. Baseband modulation

The main baseband modulation techniques considered in OWC include (i) pulse amplitude modulation (PAM), (ii) pulse position modulation (PPM), (iii) pulse interval (PIM) modulation and (iv) carrierless amplitude phase (CAP) modulation.

i. PAM

PAM is a simple modulation technique which is widely considered in OWC systems. On-off keying (OOK) modulation is a type of PAM with only two levels. OOK is a suitable OWC modulation approach and is easy to implement in such systems due to the ability of LDs and LEDs to switch on and off quickly. However, OOK is not efficient when compared with other modulation formats [14].

ii. PPM

PPM is another type of modulations that is widely considered in OWC systems [17]. Differential PPM (DPPM) was evaluated in OWC system to increase the achievable data rates. It consumes less power on average when compared to PPM. However, the distortion in the DPPM signal is greater than in the PPM signal [30].

iii. PIM

PIM encodes the information by inserting empty slots between two pulses. Its design relies on accurate synchronization which can increase its complexity compared to PPM. Digital pulse interval modulation (DPIM) is a digital version of PIM. The performance is improved in DPIM compared to PPM by removing the redundant frame space in PPM, however errors can propagate from a frame to the next [31].

iv. CAP

CAP modulation is a promising modulation format that can be used in OWC systems to achieve higher data rates. It is a bandwidth-efficient two-dimensional passband transmission scheme [14]. In CAP modulation, two orthogonal filters are chosen to modulate two different data streams.

OOK and PPM are the most popular modulation schemes applied in OWC systems [32].

b. Multicarrier modulation

More advanced techniques can be used in OWC systems to transmit multiple carriers, such as subcarrier modulation (SCM). This can provide multiple access for simultaneous users and a high data rate. However, SCM is not power efficient like single carrier schemes [33]. For instance, each quadrature phase shift keying (QPSK) or binary PSK subcarrier requires about 1.5 dB more power than OOK. In [34] a high data rate was achieved while reducing the average power requirements in SCM modulation. In [35], orthogonal frequency division multiplexing (OFDM) was applied in indoor OWC systems to achieve high data rates over a noisy channel and to reduce ISI. The system, however, did not achieve a high SNR. The main disadvantages of OFDM are the sensitivity to frequency offset and phase noise as well as the high peak to average power ratio (PAR) [36]. In general, the use of complex modulation leads to an improvement in the performance of the OWC systems, such as mitigating the ISI effect and increasing the data rates. However, these modulation techniques require a complex transceiver.

c. Multicolor modulations

Multicolor modulation has recently been considered to provide high data rates or multiple access for users [11], [14]. White light can be generated from red, green and blue (RGB) type LEDs which means data can be transferred through each color or wavelength. Wavelength division multiplexing (WDM) is the multiplexing used here. In addition, [37] achieved white light from four color LDs which provide a better result compered to three colors LEDs in term of multiple access and higher data rates due to the improved modulation capabilities of LDs.

III. Transmitter Architectures in OWC Systems

This section presents a review of the available transmitter architectures in OWC systems. The CDS system, the most commonly investigated transmitter - receiver configuration, is widely considered [4], [9], [21], [23]. A number of spot diffusing transmitter architectures were proposed and evaluated. A uniform multi-spot diffusing transmitter system was first proposed in [38]. While a diamond multi-spot diffusing system configuration can be found in [39]. In addition, there are two attractive methods for multi spot diffusing configurations: the LSMS and the beam clustering method (BCM) proposed in [23], [39], [40]. The last two methods have also been widely investigated by other researchers [41] – [47]. The results of these investigations are very promising. Different transmitter structures for IRC are shown in Fig. 7. Table 3 summaries the main advantages and disadvantages of the different transmitter structures in OWC indoor systems.

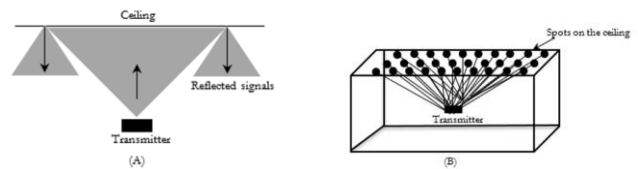

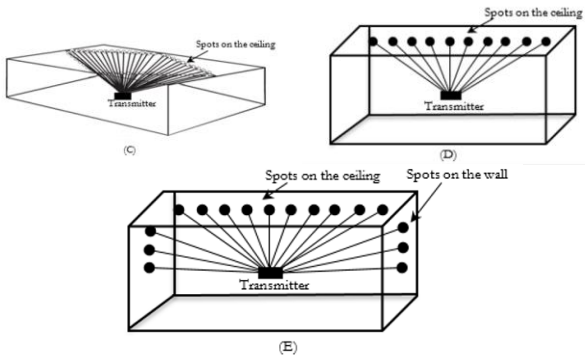

Fig. 7: Different spot diffusing transmitter structures in IRC systems; (A) Spot diffusing transmitter [4], (B) uniform multi-spot diffusing transmitter [47], (C) diamond multi-spot diffusing transmitter [38], (D) line strip multi-beam transmitter [36], (E) multi-beam clustering transmitter [39].

Table 3: Comparison of different transmitter structures in IRC indoor systems

| Type of Transmitter | Advantages | Disadvantages |
|---|---|---|
| **Spot Diffusing transmitter for CDS system [4].** | Provides link robustness against shadowing and blockage. | LOS links do not exist. (Signal is received due to reflections from walls and ceiling). High path loss. High delay spread. ISI. |
| **Uniform multi-spot diffusing [38].** | Enhanced received power, due to existence of LOS link between spots and receiver. Good coverage area. | High delay spread due to large numbers of beams. ISI. |
| **Diamond multi-spot transmitter [39].** | Provides good coverage of the sides and corners of the room. | Signal power received is very low in the room center. High delay spread. |
| **Line strip multi-beam transmitter (LSMS) [23].** | Significant improvement in the received power compared to the techniques above. Reduces ISI when combined with angle diversity receivers. | LSMS suffers from delay spread when a wide FOV receiver is employed (see Fig. 4A). Power received in the sides and corners of the room is very low. Does not provide full mobility in the room. |
| **Multi-beam clustering transmitter (BCM) [39].** | Provides full mobility for the optical receiver. Increased power received at the sides and the corners of the room (good coverage of the surroundings and room center). | The delay spread is still high when using wide FOV receiver (see Fig. 4A.). |
| **Adaptive Line strip multi-beam transmitter (ALSMS)[28], [48], [43]** | Improved allocation of power to spots according to receiver location | Increased complexity to achieve optimum transmitter power allocation |
| | Reduced delay spread and increased SNR | Single line of spots, hence poorer coverage in some areas |
| **Multi beam transmitter with Power and Angle Adaptation System (MBPAAS) [44]** | Improved coverage through beam steering and optimum allocation of power to spots. Further reduction in delay spread and increase in SNR | Increased complexity as the optimum beam direction has to be determined and the optimum power allocation has to used |
| **Beam Delay and Power Adaptation LSMS transmitter (BDAPA-LSMS) [41]** | Adapting the differential delay between the beams leads to transmitter pre-distortion and hence improved data rates | Complex technique, but simple implementation algorithms given |
| **Beam steering transmitters [12]** | Improved LOS component, hence reduced delay spread and improved SNR | User mobility is difficult |

## IV. RECEIVER STRUCTURES USED IN OWC SYSTEMS

There are many types of receiver configurations for use in OWC systems. The single element receiver with a wide FOV (90°) is the most basic configuration. The use of a wide FOV receiver enables improved collection of the optical signal. On the other hand, huge amounts of undesirable ambient light are also received with the desired optical signal and this reduces the SNR. An angle diversity receiver (ADR) that employs narrow FOV elements aimed in different directions is an attractive receiver architecture. Such ADRs were proposed in [49], [50] however the same FOV was used for all branches in the ADR.

The authors in [21], [23] developed and optimized ADR structures using an optimum FOV in each receiver branch. They also considered several diversity schemes, such as selection combining (SC), maximum ratio combining (MRC) and equal gain combining (EGC). The MRC approach can achieve better performance compared to the other methods [39]. An ADR has several advantages: reduced effect of ambient light noise, reduced signal spread and improved SNR. These advantages are realized as MRC does not ignore the unselected branches (as SC does) where these branches can make a useful contribution. It also does not simply give an equal weight to all receiver branches (EGC case) with no regard to their SNR. Its main drawbacks are design complexity, high cost (separate concentrator used for each detector) and bulky size.

Significant performance improvements can be achieved through an imaging receiver, which is one of the most attractive receivers in OWC systems due to its reduced complexity compared to ADR (only one image concentrator (eg. lens) is shared among all pixels), reduced transmitter power needed, reduced signal spread hence reduced ISI and therefore support for high data rates. It also allows space division multiple access (SDMA) to be used as the image formed resolves the signals received from different directions. An imaging receiver was originally proposed in [51]. In [51] the imaging receiver improved the SNR by 20 dB. Development and modification of the imaging receiver can be found in [41], [28], [43]–[45], [52],[10]. where the impact of the number of pixels was considered together with change in the shape of pixels, use of multiple imaging receivers mounted in different directions and power adaptation of transmitter power to fit the imaging receiver. These changes enhanced the performance of the overall system. Fig. 8 illustrates the main types of receivers. An adaptive receiver was presented in [53] for reducing ISI.

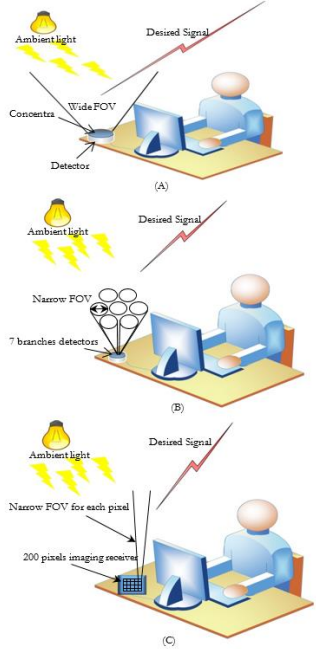

Fig. 8: Types of receivers used in IRC systems; (A) single element wide FOV ($90^0$) receiver, (B) angle diversity receiver employing narrow FOV elements and (C) imaging receiver using single imaging concentrator.

V. MAIN IMPAIRMENTS IN OW SYSTEMS

OWC systems including IRC and VLC face many challenges. In this section we will cover the main impairments in IRC systems. While, the VLC system impairments will be covered in Section VI.

The major design limitations encountered in IRC systems include: background noise, multipath dispersion, transmission power restrictions and the large capacitance associated with large area photodetectors.

• Background noise

OWC receivers detect the desired signal as well as background noise due to ambient light. The main sources of ambient light in OWC systems are sun light, fluorescent and incandescent bulbs [42]. The substantial amount of power emitted by these sources within these wavelength ranges is detected in the OWC system. This can introduce shot noise at the receiver as well as saturate the photodetector when their intensity is high [16]. The impact of ambient light can however be reduced by using optical and electronic filters. Inexpensive daylight filters can eliminate the components of the sunlight below 760 nm [54]. Fig. 9 shows the spectrum of the different sources of ambient light and the responsivity and transmission of a typical silicon p-i-n photodiode that employs a daylight filter. The characteristics and effects of ambient noise on the OWC system transmission have been investigated widely in [16], [17], [53] – [56]. The work has found that the background noise contribution from fluorescent light is negligible compared to incandescent and sunlight light (see Fig. 9A). Fluorescent lights produce most of their optical power outside the pass band of commonly used filters [16], [55].

Various techniques have been proposed and investigated to eliminate the effects of *directional* ambient light noise. For instance directional ambient noise can be reduced through the use of the fly eye receiver proposed in [38], which can optimally select a reception direction where ambient noise (for example from a window or spot light) is minimum [7], [58]. An imaging receiver can introduce better spatial resolution of signals and directional noise [41], [28], [43]–[45], [52],[10]. The authors in [21] proposed a pyramidal fly-eye diversity antenna to reduce the impact of background noise. However, even with its great advantages, this design still needs more optimization in terms of azimuth and elevation angles, triangular base dimensions as well as number of sides. In [59], an ADR was used to mitigate the background noise. The main problem with this receiver is the size of the optical elements. In addition, the imaging receiver was investigated widely, and one of its main characteristics is reduced background noise [41], [43], [51], [60].

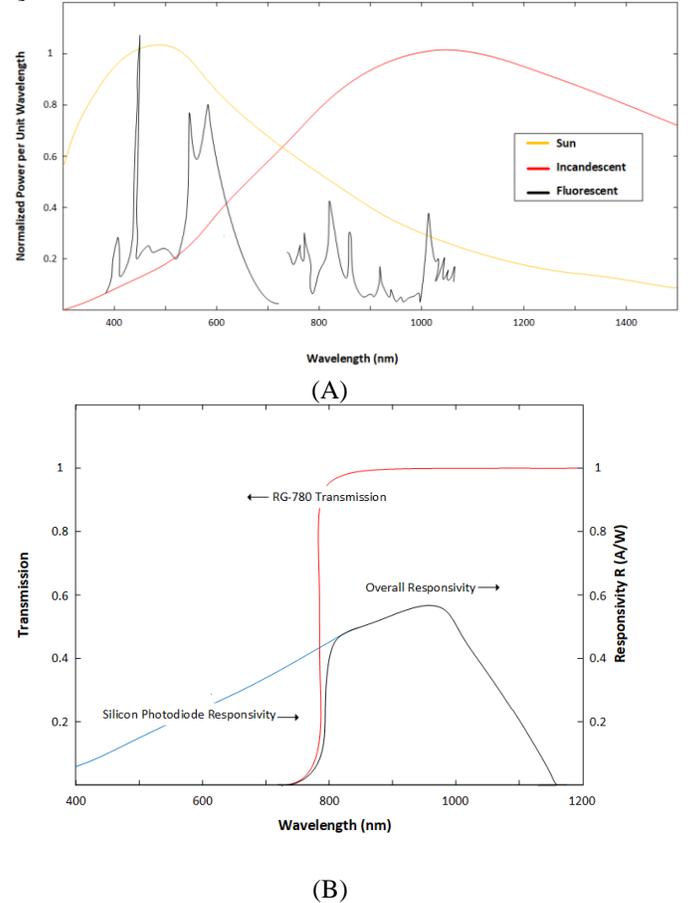

Fig. 9: (A) Power spectra of the main sources of ambient light, (B) responsivity and transmission of a silicon p-i-n photodiode employing daylight filter [22].

• Multipath dispersion

The signal received through multipath channels may lead to ISI. ISI is normally characterized through the delay spread of the channel. Multipath dispersion and ISI have a huge impact on the performance of the OWC system. In [21], the optimum value of the FOV (60º) was determined for an ADR in the presence of ISI and background noise. A considerable decrease in multipath induced dispersion can be achieved through careful design of the detecting element's (orientation and their) directionality. However, the main drawbacks of such a receiver (i.e. ADR) are their bulky size and high cost of fabrication. An imaging receiver is a suitable alternative and can be effective in combating multipath dispersion and ISI [41]. Beam angle and delay adaptation techniques were proposed by the authors in [42], [61] to improve the link performance and reduce the mobility induced impairments.

- Transmission power restriction due to eye safety

Transmitted IR beams, if operated incorrectly, can cause injury to eyes and/or skin. However, the damage to the eye is more significant due to the ability of the eye to concentrate and focus optical energy. The eye can focus light to the retina within the wavelengths 400-1400 nm [62]. The cornea (front part of eye) absorbs other wavelengths before the energy is focused. There are a number of factors that determine the level of the radiation hazard, for example, beam characteristics, operating wavelength, exposure level, exposure time and distance from the eye [18]. The optical power generated by LDs and high-radiance LEDs within the wavelength band 700-1550 nm may cause eye damage if absorbed by the retina. The maximum permissible exposure (MPE) levels are very small in this band because the incident light can be focused by the human eye onto the retina by a factor of 100,000 or more [63]. According to the international electro technical commission (IEC) standards [64], an IRC system that employs a laser source must not transmit power in excess of 0.5 mW (class A) [64]. Different transmitter technologies are used to meet the IEC standard when using LDs with high transmit power. Safety regulations have been revised downward by a factor of 100 when using a hologram [65], (typically computed by computer driven algorithms and known as a computer generated hologram (CGH)). In addition, using an optimum holographic diffuser leads to reduced path loss for the diffused signal [66]. An alternate solution can be found in [67] where an integrating-sphere diffuser for an IRC system, was designed. This diffuser is eye safe up to 0.84 W at 900 nm.

Skin safety should be considered in IRC system design. Short term effects, such as skin heating, are considered in eye safety regulations, due to the power level regulations for eyes allowing lower power levels than the power levels allowed for skin [68].

- Photodetector capacitance

Many OWC systems employ intensity modulation and direct detection. The SNR of the DD receiver is proportional to the square of the received optical power. Furthermore, the transmitted power is limited by eye safety regulations and power consumption. These constraints mean that a photodetector with a large photosensitive area should be used to gather maximum power, to improve the SNR. Unfortunately, a large photosensitive area leads to high capacitance (as the photodetector capacitance is proportional to the area of the photodetector). Large detector capacitance limits the available receiver bandwidth due to the capacitance at the input of an amplifier acting as a low pass filter. The authors in [69] proposed the use of an array of photodetectors instead of a single photodetector to mitigate the effects of large capacitance and maximize the collected power at the same time. The photodetector's effective area can be enhanced by using a hemispherical lens as suggested in [22]. Bootstrapping was proposed by authors in [27] to minimize the effective capacitance of the large area photodetector.

## VI. VLC Systems

In the past few years, the green agenda has made a huge impact on the field of communications [70]–[79]. This has pushed many researchers to produce and investigate reliable and energy efficient communication systems [80]–[91]. VLC systems are energy efficient communication systems as communications is a useful by-product of the lighting / illumination system. The main factors that fueled the growth in VLC systems are the recent developments in solid state lighting (SSL), which already provides commercial high brightness LEDs of about 100 lx - 200 lx [92], [93], with a longer lifetime (about six years) in comparison to conventional artificial light sources, such as incandescent light bulbs (whose lifetime is about four months, continuous). In addition, SSL has fast response time (modulation rate), smaller size, low power consumption and no known health hazards. Fig. 10 shows the practical luminous efficacy of different available light sources [92].

The dual functionality of the VLC system, i.e. illumination and communication, makes it a very attractive technology for many applications, such as car-to-car communication via LEDs, lighting infrastructures in buildings for high speed data communication, traffic light management in trains and high data rate communication in airplane cabins. The first proposed VLC system was at Keio University in cooperation with Nakagawa laboratory in Japan [70] – [72].

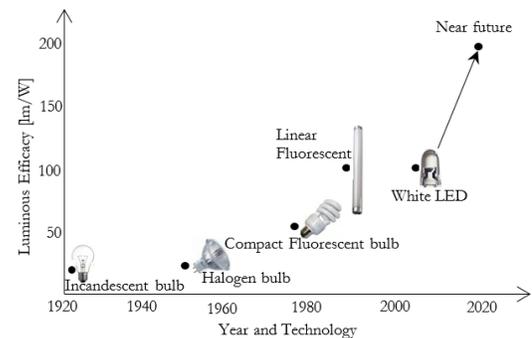

Fig. 10: Comparison of different light sources in terms of luminous efficacy [92].

### A. VLC Standards

In 2003, VLC system standardization was started in Japan [97]. The Japan Electronics and Information Technology Industries Association (JEITA) proposed two standards based on visible light communication in 2007. The first one is the visible light communications system standard (CP-1221) while the second is the visible light ID system standard (CP-1222) [98]. Six years later, a new Japanese standard based on CP-1222 was proposed, the visible light beacon system standard (CP-1223).

In 2007, the Smart Lighting Engineering Research Centre (ERC) was established in the USA funded by the National

Science Foundation (NSF) for developing solid state lighting systems that have an advanced functionality [99].

In 2011, the IEEE published the physical and medium access control (MAC) layers as a standard for VLC systems, called IEEE 802.15.7 [100]. This standard can provide a data rate of up to 96 Mbps for transmitting video and audio services. The standard also considers the mobility of visible light communication links as well as noise and other interference sources. The IEEE 802.15.7 standard utilizes only the visible light spectrum and considers optical receiver specifications. In the same year, a technology called light fidelity (Li-Fi), which has features similar to the Wi-Fi technology but uses light as transmission medium instead of RF, was launched [101]. Three years later, the IEEE 802.15.7 standard group revised the standard by releasing an update, the IEEE 802.15.7m. The updated standard utilizes infrared, visible light and near ultraviolet wavelengths as medium, also, considering optical camera communications (OCC) and imaging receivers as detectors [102]. In 2017, the IEEE 802.15.13 standard was published for providing data rates up to 10 Gbps with range up to around 200 m [103]. The standard is designed to support point-to-point and point-to-multipoint communications that can be in both coordinated and non-coordinated topologies. One year later, in 2018, the IEEE announced a new light communications task group for devolving a global standard in wireless local area networks based on visible light communication. The standard, named IEEE 802.11bb, intends to engage with manufacturers, operators and end users to build the system [102].

### B. VLC sources

The VLC system structure is similar to that of IRC systems. However, the main essential difference between VLC and IRC systems is the function of the transmitter. The transmitter in VLC systems is used as an illumination source as well as a communications transmitter for data users. Therefore, the VLC network designers must consider the following two requirements. Firstly, the visible light transmitter (LEDs or LDs) should work as sources of illumination. The optimum lighting configuration and LED angles have been studied in previous work [104], [105]. According to the international standards organization (ISO) and European standards, the illumination should be 300 lx -1500 lx for an indoor office [106]. Secondly, the data should be transmitted through LEDs without affecting the illumination.

An essential advantage of VLC systems is that they can provide communication and illumination at the same time. White light can be generated by using LEDs or LDs. The following sections describe each source.

1. LEDs

At present, there are two different approaches generally used to generate white light from LEDs. In the first method white light is generated using a phosphor layer (emitting yellow light) that is coated on blue LEDs. The light emitted by blue LEDs ($\lambda \approx 470$ nm) is absorbed by the phosphor layer, where the blue wavelength excites the phosphor causing it to glow white, then the phosphor emits light at longer wavelengths. The second method uses RGB LEDs. With the correct mixing of the three colours, red, green and blue, white light can be created such as in colour TV. The lower cost and complexity of the first method makes it more popular. However, the modulation bandwidth is limited to tens of MHz due to the slow response of the phosphor which limits the communication data rate. Both techniques have been demonstrated in conjunction with VLC [107], [108]. Fig. 11 depicts the methods used to generate white light from the LEDs.

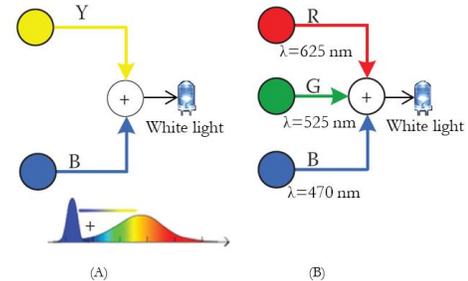

Fig. 11: Two approaches used to produce white emissions from LEDs, (A) phosphor LED method, (B) RGB method.

2. LDs

Laser diodes (LDs) can be used as sources for illumination and communication in VLC systems. The experiment in [37] shows that mixing different LD colors can provide a white light source. Fig. 12 shows the experimental setup used to generate white color using 4 different LDs colors. The four laser colors are combined using chromatic beam-splitters. The combined laser colors reflected by a mirror pass through a ground-glass diffuser which reduces speckle before the illumination.

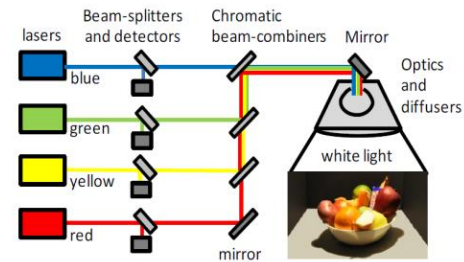

Fig. 12: Generating white light using 4 different LD colors [37]

A simple physical layer block diagram of a VLC system is shown in Fig. 13. Digital and analogue components are used in the transmitter and the receiver. The data stream, baseband modulator and digital to analogue convertor (DAC) are the digital components in the transmitter. Similarly, the digital components in the receiver are the analogue to digital converter (ADC), demodulator and data sink. The trans-conductance amplifier (TCA), bias Tee and LEDs are the analogue components in the transmitter. The receiver includes the photodiode (PD), trans-impedance amplifier (TIA) and band pass filter (BP). The AC signal in the transmitter is added onto the DC current by a bias Tee (a device used in a high frequency transmission line to produce a DC offset). Since LEDs work in a region with unipolar driving currents, the absolute driving current (AC+DC) has to be larger than zero. The total current is fed to the LEDs to emit the modulated output power. The power received by the PD is converted into a current (I-PD), and this current consists of two components, the AC and DC parts. The AC component is amplified and then filtered using TIA and BP respectively. Finally, the digital signal is demodulated after conversion using ADC.

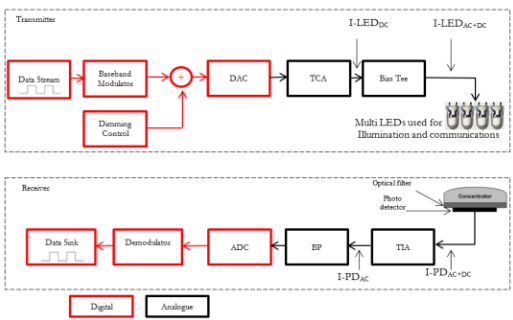

Fig. 13: Block diagram of VLC system.

Brightness (light dimming) control for the LEDs is required in VLC systems as these sources are also used for illumination. Many methods have been proposed to implement dimming control [109]–[112]. The simplest method is amplitude modulation (AM) dimming: the luminous flux is controlled by controlling the input DC current. However, the chromaticity coordinates of the emitted light can change [113]. Pulse width modulation (PWM) is another method that controls the width of the current pulse. The main feature of PWM dimming is that the amplitude of the pulse is constant. The width of the pulse varies according to the dimming level, thus resulting in the emitted light spectrum being constant. Binary code modulation or bit angle modulation (BAM) is a good dimming method that was introduced by artistic licence engineering (ALE). It is a new LED drive technique that can be used in VLC systems [114]. The main advantages in using this technique are implementation simplicity, the potential to reduce flicker compared to PWM, lower processing power and simple data recovery operations.

BAM encoding for LED dimming makes use of the binary data pattern shown in Fig. 14. In the 8 bit BAM system, the pulse width of bit 7 (most significant bit, (MSB)) is equal to $2^7=128$, whereas the pulse width of bit 0 (least significant bit, LSB) is equal to $2^0=1$ (a unit width). This allows continuous brightness levels to be defined where the granularity of the dimming levels depends on the number of bits used in BAM.

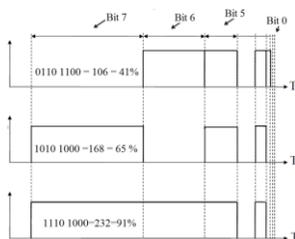

Fig. 14: BAM signal waveform used in dimming control.

As mentioned earlier, the primary function of LEDs is to provide illumination, so it is essential to define the luminous intensity, the quantity used to determine the illumination level. To get comfortable office lighting, a certain amount of luminance is required (300 lx -1500 lx) [106]. The brightness of the LED is expressed as luminous intensity. Luminous intensity is defined as the luminous flux per solid angle as [96]:

$$I = \frac{d\emptyset}{d\Omega} \tag{12}$$

where $\Omega$ is the spatial angle and $\emptyset$ is the luminous flux. From the energy flux (emission spectrum), $\emptyset e$ and $\emptyset$ can be calculated as [96]:

$$\emptyset = F \int_{380}^{780} A(\lambda)\, \emptyset e\,(\lambda)\, d\lambda \tag{13}$$

where $A(\lambda)$ is the eye sensitivity function and $F$ is the maximum visibility, which is equal to 683 lumen/Watt (lm/W) at 555 nm wavelength. Assuming that the LED light has a Lambertian radiation pattern, the direct luminance at any given point in an office can be estimated as:

$$E = I(0) \frac{\cos^n(\theta)\cos(\gamma)}{D^2} \tag{14}$$

where $I(0)$ is the centre luminous intensity of the LED measured in candela (cd), $\theta$ is the irradiance angle, $D$ is the distance between the LED and the point of interest, $\gamma$ is the angle of incidence and $n$ is the Lambertian emission order, defined as [22]:

$$n = -\frac{\ln(2)}{\ln\left(\cos\left(\Phi_{\frac{1}{2}}\right)\right)} \tag{15}$$

where $\Phi_{\frac{1}{2}}$ is the semi angle at half power of the LED.

Using the relations above, the illuminance distribution is determined in a room with dimensions 5m×5m×3m, where four LED light units with specifications similar to those in [96] were used. Fig. 15 illustrates the illumination distribution. Note that the illumination is within the acceptable range of (300 lx -1500 lx).

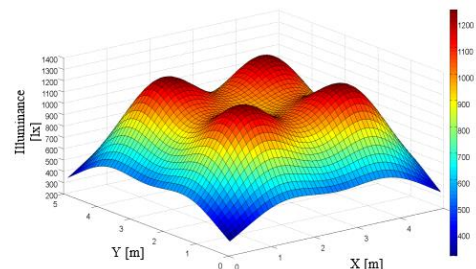

Fig. 15: The distribution of illuminance, Min. 315 lx, Max. 1220 lx.

C. Design challenges in VLC systems

VLC systems are a potential candidate to achieve multi gigabits per second data rates in indoor wireless systems. There are however several impairments that VLC systems have to deal with. These impairments include the ISI caused by multipath propagation, the low modulation bandwidth of the transmitter (i.e. LEDs), and the provision of an uplink connection for VLC systems.

1. ISI in VLC systems

In real environments, there is more than one transmitter (LEDs), and these transmitters are used for illumination as well as communication. The VLC receiver therefore collects signals from more than one transmitter, and each signal arrives using potentially a different path with a different delay profile. In addition, due to the reflections from the walls, ceiling and other objects, the received optical signal suffers dispersion due to multipath propagation. This results in ISI, especially at high data rates. In the literature, many techniques have been proposed to mitigate ISI in VLC systems. Among the most notable solutions is the use of return-to-zero OOK (RZ-OOK) modulation with the space between pulses being used as a guard interval to reduce the effect of the delay spread in low data rate applications [115]; while, in [116] the authors used OFDM to reduce the ISI. Spread spectrum has also been considered to combat ISI, however, it reduces the bandwidth efficiency as well [117]. The authors in [118] used zero forcing (ZF) equalization with transmitter (i.e. LED) arrangements to reduce the effects of ISI. The bit error rate (BER) achieved using this technique was similar to the channel without ISI. In [96] it was found that decreasing the receiver field of view leads to reduced ISI, whereas increasing the data rate leads to an increase in ISI. The use of adaptive equalization using the least mean square algorithm has the ability to reduce the effects of ISI and was investigated at data rates up to 1 Gbps [119]. The authors in [120] used MIMO approaches to reduce shadowing effects. Other approaches can be used to reduce ISI in VLC systems. For instance, using pre and post equalization techniques, coding, angle diversity receivers, and imaging receivers [10]–[13].

2. Low modulation bandwidth of the LEDs

The modulation bandwidth of the transmitters (i.e. LEDs) is typically less than the VLC channel bandwidth, which means that the former limits the VLC data rates.

To achieve high data rates in VLC systems a number of different techniques can be used, for instance, optical filters, pre and post equalization (or both), complex modulation techniques and parallel communication (optical MIMO, (OMIMO)). A blue optical filter at the receiver can be used to filter the slow response yellowish component, and this approach is probably the simplest and most cost effective approach to increase the VLC data rates [121]. However, the achieved bandwidth is insignificant (8 MHz). The drop in the white LED response can be compensated for by using a simple analogue pre-equalizer at the transmitter, and this technique was shown to offer 40 Mbps without the use of a blue filter [122]. However, the data rate achieved is still very low compared to the available VLC spectrum. A moderate data rate (80 Mbps) can be achieved using more complex pre-equalization [123]. Pre-equalization however has the drawback that the drive circuit of the LED needs to be modified, and this leads to higher costs and lower emitter efficiency (i.e. not all the input power is converted to light). By combining simple pre and post equalization, 75 Mbps can be achieved [124]. Both analogue and digital equalization techniques can be applied. Analogue equalization techniques are suitable for OOK modulation, while digital techniques are preferable for OFDM [125]. Using least mean square (LMS) and decision feedback (DFE) equalizers at data rates above 200 Mbps was shown to be very effective to mitigate ISI [119]. The use of such equalizers however leads to a computationally complex receiver structure. A data rate of 100 Mbps - 230 Mbps was the maximum data rate achieved using phosphorescent LEDs with simple OOK modulation [126]. The modulation scheme used is very important in VLC systems because it affects the achievable data rate, BER and luminance. For example, multiple PPM was used for modulation and dimming control [111]. Different types of modulation were evaluated in terms of their combined modulation and dimming control capabilities. The results showed that overlapping PPM (OPPM) can achieve flexible luminance and good performance [127]. Higher data rates can be achieved when complex modulation approaches are used. For example, Discrete Multi-tone modulation (DMT) can provide data rates of about 1 Gbps [108]. This type of modulation requires however a complex transceiver. In terms of parallel transmission, VLC systems have naturally favorable features for OMIMO as many LEDs (transmitters) are used for illumination and different data streams can be sent on every single LED to maximize the throughput. At the same time, an array of photodetectors can be used at the receiver. This setup offers improvements in security, link range and data rates while the power required is unaltered [128], [129]. Detailed study of OMIMO revealed that channel cross talk is a critical limiting factor in VLC systems [130]. A significant enhancement in the data rates can be achieved with RGB LEDs. A 1.25 Gbps system was reported in [131] using RGB LEDs in a single color transmission mode, and 1.5 Gbps was achieved using a new micro LED (µLED) array with non-return-to-zero OOK (NRZ-OOK) as the modulation scheme. The 3 dB modulation bandwidth of this LED was 150 MHz [132]. The maximum data rate achieved using commercial RGB LEDs with low complexity modulation (OOK) was 500 Mbps [133]. Recently, a 3 Gbps VLC system based on a single µLED with OFDM modulation was successfully demonstrated [134]. The highest throughput achieved by LEDs was reported in [107], which reported an aggregate throughput of 3.4 Gbps using DMT, wavelength division multiplexing (WDM) and RGB LEDs. The design and implementation complexity are a major concern in these systems however.

3. Provision of an uplink connection in VLC systems

Using white illumination LEDs for communication is naturally one directional communication (downlink). So, providing an uplink connection can be a major challenge. Many techniques have been proposed to provide an uplink for VLC systems. In [135], the use of retro reflecting transceivers with a corner cube modulator to provide low data rate uplinks was proposed for visible light communication as shown in Fig. 16A. In [136], [137] an IR transmitter was proposed to provide an uplink for VLC systems. In [138] the authors suggested that RF systems can be employed in conjunction with the VLC system to provide reliable bidirectional communication as shown in Fig. 16B. However, all the above techniques add complexity to the VLC systems and do not provide a high data rate uplink connection. Recent research demonstrated high data rate bi-directional VLC transmission with 575 Mbps downlink and 225 Mbps uplink using SCM with WDM [139]. In addition, other recent research proposed a solution for the uplink VLC system using IRC. As a result, the uplink data rate reached 2.5 Gbps [140].

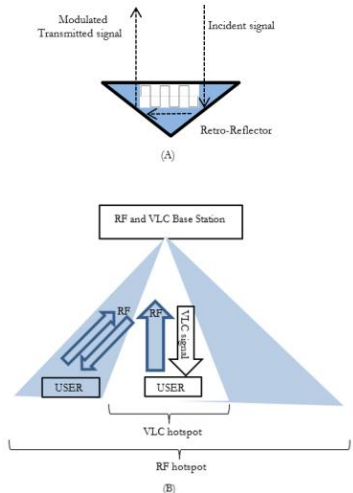

Fig. 16: Two techniques used to provide an up-link connection in VLC systems, (A) retro reflecting transceivers and (B) bidirectional VLC system combined with an RF system.

### D. VLC system architectures

Popular wireless system architectures used in RF wireless systems have also been explored in VLC systems. Fig. 17 illustrates four system architectures: single input single output (SISO), single input multiple output (SIMO), multiple input single output (MISO) and multiple input multiple output (MIMO). The number of inputs and outputs in RF communication systems is dictated by the number of antennas at the sender and the receiver, while, in VLC systems it is based on the number of sources (LEDs, or LDs) and the number of optical receivers.

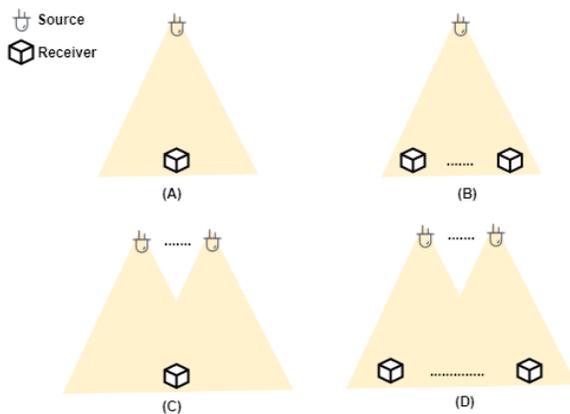

Fig. 17: VLC system architectures (A) SISO (B) SIMO (C) MISO and (D) MIMO.

MIMO approaches were widely investigated in VLC systems to provide high data rates and to support multiple users [117] – [122]. Imaging and non-imaging receivers are investigated in VLC systems [141]. A hybrid diffusing IR transmitter was studied to support VLC systems when the light is turned off or dimmed [52]. In addition, the authors in [142] developed MIMO VLC systems by combining transceivers and dimming control.

### E. Multiplexing techniques

OFDM and WDM are two of the most widely considered multiplexing techniques in VLC systems. Their main features are summarised here.

#### 1. OFDM

The use of OFDM in VLC systems offers the ability to combat ISI. High data rates can be achieved by dividing the available bandwidth into sets of orthogonal subcarriers [147]. However, OFDM signals suffer in-band and out-band distortion due to the high peak to average power ratio (PAPR). This flicker in the power leads to reduced device lifetime and eye safety problems [148].

Numerous approaches have been proposed to realize optical OFDM (O-OFDM), ie approaches that satisfy the IM/DD constraints. These include direct-current (DC) O-OFDM (DCO-OFDM) [36], asymmetrically clipped (AC) O-OFDM (ACO-OFDM) [149], Flip-OFDM [150] and Hermitian symmetry (HS) free (HSF)-OFDM (HSF-OFDM) [151].

#### 2. WDM

WDM allows the multiplexing of different data sources on different wavelengths and then into a single channel. It promises great flexibility and bandwidth efficiency, as it did in the case of fibre optic communication systems. It is feasible to apply WDM in VLC systems, as white light is composed of a combination of colors which can be produced using colored LEDs (RGB) or colored LDs (BGYR). Colored LEDs can be individually modulated and an optical multiplexer then transmits the parallel data. At the receiver the signal is passed through a de-multiplexer (optical filters used as de-multiplexer) and then optical receivers are used (each optical receiver thus detects a certain wavelength) for data recovery. Fig. 18 shows a WDM VLC system. By using LEDs (RGB), a WDM data rate of 3.22 Gbps was reported in [152]. Another demonstration was reported in [153] of a VLC system that employs WDM. The use of LDs (BGYR), in VLC systems was proposed in [11] resulting in a data rate of 10 Gbps.

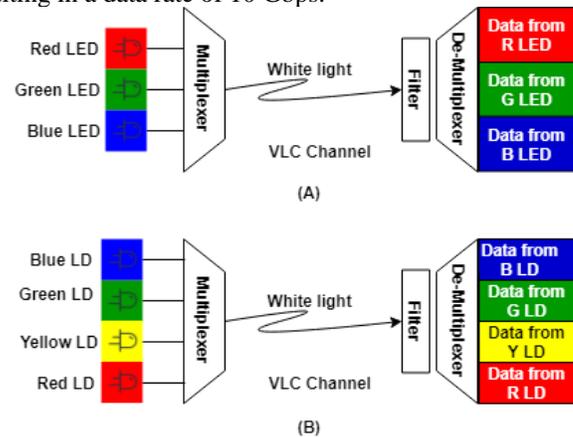

Fig. 18: WDM used in VLC systems (A) using LEDs and (B) using LDs.

The effect of the change in the correlated colour temperature due to modulation has not been substantially explored. An important feature of the source of light is its apparent colour as a perfectly white source when viewed. This feature can be quantified using chromaticity coordinates on a chromaticity diagram. The International Commission on Illumination (CIE) created in 1931 the RGB colour space diagram shown in Fig. 19. The circumference of the area includes the coordinates of all the real colours. A black body radiator at different temperatures produces different colours. The white colour is defined by the region close to the black body radiators locus (starting at approximately 2,500 K). RGB hues are located within regions that span from the white region towards the corresponding corners of the figure. Sources that are very close to the Planckian locus may be described by the colour temperature (CT) [154].

The chromaticity coordinates can be defined as [154]:

$$X = \frac{\sum_{i=1}^{n} x_i \emptyset_i}{\sum_{i=1}^{n} \emptyset_i} \qquad (16)$$

and

$$Y = \frac{\sum_{i=1}^{n} y_i \emptyset_i}{\sum_{i=1}^{n} \emptyset_i} \qquad (17)$$

where $\emptyset_i$ is the radiant flux and $x_i$ and $y_i$ are the primary sources' coordinates. From Fig. 19, a white colour can be obtained from two colour sources (blue and yellow, see dotted line in Fig. 19). In addition, a white colour can be obtained from three sources (see solid triangle in Fig. 19).

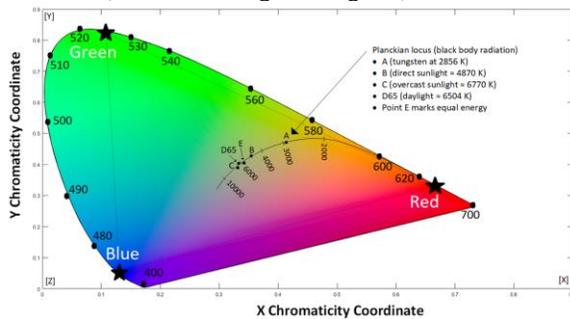

Fig. 19: 1931 CIE chromaticity diagram [154].

F. Multiple access schemes

Several multiple access (MA) schemes have been studied for use in VLC systems. Some of the MA schemes that have been used in RF systems are also used in VLC systems. These include time division multiple access (TDMA), frequency division multiple access (FDMA), space division multiple access (SDMA), code division multiple access (CDMA) and non-orthogonal multiple access (NOMA).

1. TDMA

TDMA for multi-user VLC systems was proposed and investigated in [155]. It is also sometimes used for preventing collisions where other MA schemes may fail, such as situations where two transmitters have the same distance to the receiver [29]. A controller can share time between the transmitters in a medium access control (MAC) protocol, which may also allow uneven time splitting between trasnceivers. The throughput of TDMA schemes decreases if the number of users increases. As a result the system cost may rise due to the need for synchronization between transmitters [156].

2. FDMA

FDMA is a scheme that can support multiple access based on orthogonal frequency bands for example. In VLC systems, there are four techniques based on FDMA that have been investigated [157]–[159].

The first method uses single carrier FDMA (SC-FDMA) which is based on frequency division multiplexing. SC-FDMA was proposed in [159]. It reduces the value of the peak-to-average power ratio (PAPR) and that is one of its main advantages [160].

The second technique is orthogonal frequency division multiple access (OFDMA) which was studied in [157]. OFDMA is a multi-user approach based on OFDM modulation. It is an extension of OFDM which means each user can employ a group of subcarriers in each time slot. OFDMA was evaluated in a number of studies [161]–[163]. The data rate achieved when using OFDMA can be reduced because of spectrum partitioning [164]. The third technique is orthogonal frequency division multiplexing interleave division multiple access (OFDM-IDMA), which was proposed in [157]. OFDM-IDMA is also a multi-user technique based on OFDM modulation. It is a scheme that combines OFDM and IDMA technologies. Both techniques: OFDMA and OFDM-IDMA are asymmetrically clipped at zero level after OFDM modulation. OFDM-IDMA provides good power efficiency compared to OFDMA. In addition, it was shown in [157] that signal to noise ratios (SNR) above 10 dB in a system with modulation size equal to 16, OFDM-IDMA provides better results compared to OFDMA. It is shown in [157] however that the decoding complexity and PAPR are lower in OFDMA compared to OFDM-IDMA.

The last technique is interleaved frequency division multiple access (IFDMA) which was proposed in [158]. IFDMA was introduced to reduce PAPR compared to OFDMA. In addition, the effect of the non-linear characteristics of LEDs is lower in IFDMA which can provide a higher power efficiency. The computational complexity needed is lower in IFDMA compared to OFDMA as IFDMA does not include discrete Fourier transform or inverse discrete Fourier transform operations. Moreover, it was shown in [165] that IFDMA is a promising multiple access technique in VLC due to its ability to mitigate multi-path distortion and reduce synchronization errors.

3. CDMA

CDMA is a non-orthogonal multiple access scheme which can offer high spectral efficiency compared to OFDMA and TDMA. Each user in CDMA uses a special code to provide simultaneous transmission and reception. The special code used by each user is typically an optical code; an example was proposed in [166]. In addition, random optical codes were proposed in [167], [168] to support a large number of users by spreading the signal bandwidth. A synchronization technique was reported in [169] to help prevent the undesired correlation characteristics over random optical codes. As a result, a good performance was achieved.

The authors in [170] proposed color shift keying (CSK) as a technique based on CDMA that can increase the multiple access capacity. Consequently, each transmitter can gain 3dB compared to OOK.

Multi-carrier (MC) is another approach investigated in [171], [172] based on CDMA. MC-CDMA combines features of CDMA and OFDM. MC-CDMA diffuses the data symbols of each user in the frequency domain over OFDM subcarriers.

Subsequently, the sum of the data symbols from multiple users are re-modulated in the time domain using OFDM. Additionally, the authors in [171] showed that the transmitted optical power can be reduced by using sub-carrier selection.

Another multiple access technique based on CDMA was proposed in [173] which provides interference-free links while transferring information. The technique allows users to decode their signals without pre-confirmation. However, this technique was investigated in a small area less than one square meter which reduces its practical utility.

The power allocation issue in CDMA was studied in [174]. Two types of distributed power allocation were proposed which are partial and weighted distributed power allocation, which can be optimized. The result of the study showed that the proposed techniques offer low computational complexity compared to the optimized optimal centralized power allocation approach.

Different CDMA schemes were proposed in [175] that utilize micro-LED arrays. These schemes provide a higher modulation bandwidth which can also support a large number of users.

4. SDMA

SDMA can provide multi-user support in VLC systems. It was shown in [176] that increasing the number of transmitters in SDMA VLC systems increases the throughput. Moreover, the system capacity was shown to improve more than ten times when SDMA is used compared to TDMA VLC systems. In [177] a low complexity suboptimal algorithm was introduced for coordinated multi-point VLC systems with SDMA grouping. The proposed technique offers an improvement in the system performance and fairness in throughput.

5. NOMA

NOMA was studied in [178], [179], [180] as a promising technique that can provide multiple access in VLC systems. It is also called power domain multiple access. NOMA differs from other multiple access techniques which provide orthogonal access for multiple users either in frequency, time, code or phase. Each user in NOMA can use the same band at the same time. Thus, the power level can be used to help distinguish users. In addition, the transmitter in NOMA applies superposition coding to simplify the operation at the receiver. As a result, channels are separated at the receiver for both uplink and downlink users [179]. NOMA was also studied in some recent work as a means to support multiple users at high data rates [164], [179], [181], [182].

NOMA outperforms OFDMA in terms of the achievable data rate in VLC as was shown in [179]. In addition, it was shown in [183] that the system capacity is improved when NOMA is employed. MIMO NOMA VLC systems were studied in [160]. The results indicated that this approach can offer high data rates, high capacity and higher spectral efficiency. An experimental study was reported in [164] where a MIMO NOMA VLC system was demonstrated. A normalized gain difference power allocation (NGDPA) method was proposed in the experiment to provide an efficient and low complexity power allocation approach. The experimental results showed that NOMA with NGDPA can result in up to 29% sum rate improvement compared to NOMA with the gain ratio power allocation scheme.

Table 4 provides a summary of the multiple access schemes studied in the context of VLC.

Table 2: Summary of VLC multiple access schemes

| Scheme | Summary |
|---|---|
| TDMA | TDMA can help prevent collisions between users in VLC systems. The per user throughput decreases when the number of users increases and resources can be wasted during user inactivity. Accurate synchronization is needed between the transmitters which increases cost and complexity [156]. |
| FDMA | Four main FDMA techniques have been studied in VLC systems: SC-FDMA, OFDMA, OFDM-IDMA and IFDMA. SC-FDMA reduces the PAPR. However, it may not be the best approach at high data rates and with a large number of users. OFDMA and OFDM-IDMA can provide better throughput. OFDMA offers lower decoding complexity and lower PAPR compared to OFDM-IDMA, however it can introduce spectrum partitioning which may reduce the data rate. The power efficiency in OFDMA is high compared to OFDM-IDMA. Moreover, at some SNRs and modulation sizes, OFDM-IDMA has better throughput compared to OFDMA. IFDMA can mitigate multi-path distortion and can help reduce the impact of synchronization errors. The computational complexity and PAPR in IFDMA are lower compared to their values in OFDMA. |
| CDMA | Special and random codes are utilized. Random code were proposed to increase the number of users. Synchronization techniques were added over Random codes to improve the system performance. CSK is another technique that was proposed in conjunction with CDMA. It can provide 3dB gain for each user compared to OOK. Techniques were proposed based on CDMA to provide interference-free links, but in small areas while other approaches were reported to increase the achievable data rates and number of users. |
| SDMA | SDMA is not widely studied in VLC systems to date, but is a very promising scheme. One technique based on SDMA was evaluated and shown to offer an improvement in the system performance and throughput fairness. |
| NOMA | NOMA can provide non-orthogonal multiple access in VLC systems. It can outperform OFDMA in terms of data rate and system capacity. MIMO NOMA provides further improvement in data rate, capacity and spectral efficiency [144]. |

G. Mobility in VLC systems

One of the requirements in indoor communication systems is to provide support for the expected forms of user mobility. To do this with minimal impact on the system performance adaptation may be needed in the transmitter and receiver. In IRC systems mobility can be supported through the use of adaptive LSMS, adaptive BCM, angle diversity receivers and imaging receivers [28], [43]–[45], [48] which when used together can enable the IRC system to adapt to the environment. However, adaptive techniques such as LSMS and BCM cannot be directly applied in VLC systems due to the dual functionality of the VLC source (i.e. illumination and communication). Mobility may be supported in VLC systems by using improved receivers such as angle diversity receivers or imaging receivers [10]–[13].

Beam steering approaches are already applied in FSO systems and can also be utilized in VLC systems. VLC beam steering can be achieved by using electronically controlled mirrors in front of the receiver and receiver tilting. One inexpensive approach used to provide good link quality during mobility uses mirrors with piezoelectric actuators in front of the receiver [184]. This method however, leads to a bulky receiver and cannot be used in mobile devices. Fig. 20 illustrates this technique. Another approach is the tilting of the transmitter and receiver together using piezoelectric actuators that are controlled electronically. As with the mirrors method, the tilting method also needs to be controlled using electronic circuits. The tilting approach has two different structures as shown in Fig. 21A. In the first structure all of the transceiver (LED and photodetector) are placed on actuators and in the second structure, in Fig. 21B only the lenses of the transceiver are placed on the actuators. The advantage of the second structure is the ability to zoom in and out, which means the FOV can be changed [184].

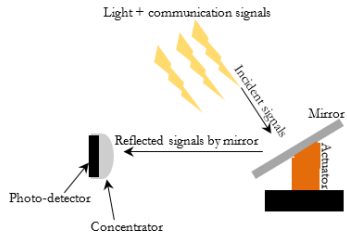

Fig. 20: Beam steering using a mirror.

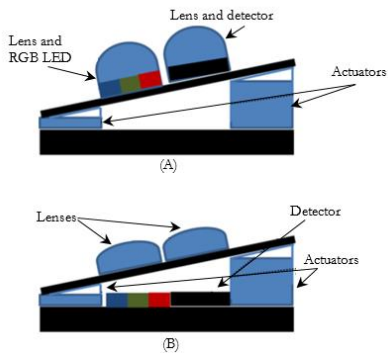

Fig. 21: Beam steering using tilting method, (A) all body placed on actuators, (B) only lenses placed on actuators.

### H. Hologram Design for VLC Systems

Holograms were recently studied in VLC systems to provide a high data rate and to improve the SNR [12], [13], [185], [186]. A hologram is a transparent or reflective device that can be used with VLC sources. It modulates the amplitude or phase of the signals that passes through it. Consequently, the light rays are diffused by splitting the light into beams to cover the required area, as shown in Fig. 22. The authors in [12] used a CGH in their study to achieve a data rate of up to 20 Gbps. In addition, fast CGH (FCGH) adaptation was proposed in [13] as a new method that can support user mobility. The data rate was also increased up to 25 Gbps, and the SNR was improved.

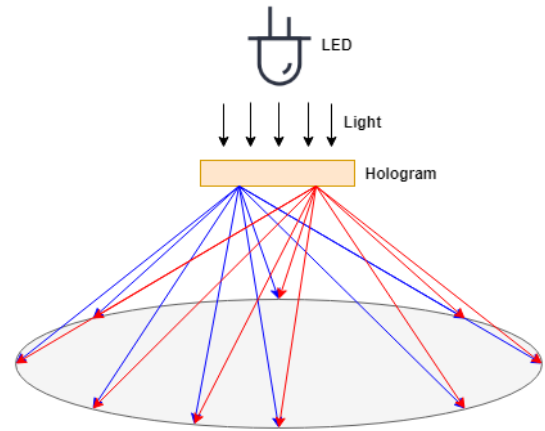

Fig. 22: Hologram and illumination produced.

### I. Security in VLC Systems

Security in VLC system can be divided into two categories, namely physical layer security and data layer security, ie encryption.

1. Physical Layer Security

VLC systems offer physical layer security for the user information, as light signals cannot travel through opaque objects such as walls. Therefore, an attacker cannot get access to the medium unless they are in the visible range for the VLC communication system. This makes VLC systems more secure than an RF communication system [29]. VLC physical layer security was studied in [163.5], [163.6].

2. Encryption

Quantum key distribution (QKD) can help secure data in OWC systems. QKD can use photons to exchange a secret key between two points. Therefore, it can increase security in OWC system by encrypting the transferred data. The secret key can be exchanged for example using a one-time pad (OTP) cryptographic algorithm as well as the characteristics of quantum physics. Eve is restricted by the quantum theory from storing or copying transferred photons via the quantum channel because of the changes in the photon characteristics. After exchanging the secret key, the data are encrypted by the secret key and transferred via classical communication. A wireless QKD approach was examined in an indoor environment [187].

### J. Future trends for VLC

VLC systems are rapidly developing, however, before VLC becomes widespread, many issues should be resolved, such as the identification of the optimum modulation techniques and the best dimming control methods, together with the optimum multiple access approaches and standardization. Currently, VLC systems do not fit with most types of LED lighting systems, as there are different drive mechanisms and architectures. So, collaboration between communications organizations and the solid state lighting (SSL) industry is necessary to establish standards for SSL systems to be used for illumination and communications [188].

## VII. Main Applications of OWC Systems

OW is generally used in three applications area: office interconnection, last mile broadband access and personal communications.

### A. Main applications of IRC systems

To date, there have been many applications that use infrared OW systems. A key application area is personal communications. OW personal communications (short range applications) are widely studied due to their low complexity, high bandwidth and low cost. Two main factors that make short range OW systems preferable from the users' point of view are the security features (light is contained in the environment where it originates) and the potentially high data rates these systems can achieve. In addition, lower transmission powers can be used and a data rate of 1 Gbps has already been standardized in the Giga IR standard with data rates up to 10 Gbps planned [189]. A range of devices incorporated IrDA ports, however these were short range and relatively low data rate. This led to the development of many subsequent standards, most notably the recent effort by the IEEE to develop an OW physical layer option in IEEE 802.11, namely the planned IEEE 802.11bb standard [190]. The IrDA standard can be considered obsolete as it was not recently updated and the maximum data rate is low. The IrDA standardized short range infrared OW for point to point communication (half duplex mode), with direct LOS being the preferred configuration for IrDA links. The IrDA link architecture does not cater for user mobility. The standard was created in 1993 by IrDA, which includes more than 160 members with the majority of these being manufacturers of hardware, peripherals, components and software [191]. This standard provides interoperability between different software and hardware devices using wireless IR. between 2004 and 2008 more than 600 million devices were produced based on IrDA technology, and this number used to increase by 20% every year according to [191], [192]. Personal communication devices that use IrDA transceivers include: printers, PDAs, electronic books, digital cameras and notebooks (most kinds of communications and computing devices using IR). Other OW indoor systems were developed, for example those that use Optical Camera Communications [193].

### B. Main applications of VLC systems

Indoor VLC applications include applications for data communications, indoor location estimation, indoor navigation for visually impaired people and accurate position measurements.

- Indoor applications

Like IRC systems, VLC systems can be used to transfer data in offices and indoor spaces. The user location can be estimated in an indoor environment using white LEDs. For example, indoor spaces can be illuminated by LEDs with a unique ID for each LED. White LEDs in conjunction with geometric sensors integrated in smart phones can help visually impaired people move inside buildings [194]. Fig. 23 shows a prototype navigation system for visually impaired people [194]. RF signals can be undesirable inside hospital environments, especially in operating theatres and magnetic resonance imaging (MRI) scanners, therefore VLC systems can be a potential solution in such scenarios.

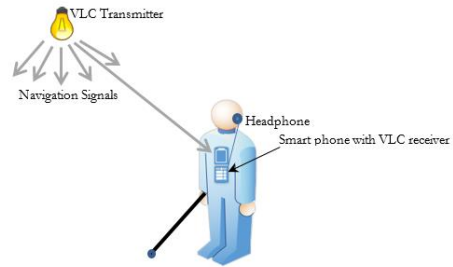

Fig. 23: Indoor VLC navigation system for visually impaired people.

Recently, various methods for VLC indoor positioning were studied [195], [196]. VLC systems are strong candidates for indoor positioning due to several features. Firstly, better positioning accuracy (few millimetres) can be achieved, which is better than the accuracy achieved by radio wave systems. The added accuracy is mainly attributed to the lower interference and multipath effects suffered by VLC systems and the short wavelength used. Secondly, VLC positioning systems can be used in environments where radio positioning systems are restricted, such as industrial settings with large metallic structures [197] or in hospitals [198].

## VIII. VLC experimental demonstrations and products

This section provides a brief overview of recent VLC projects and products.

### A. VLC experimental demonstration

1. Ultra-parallel visible light communications (UP-VLC) project

The UP-VLC project started in October 2012 and concluded in February 2017. It involved six research groups at 5 institutions and was funded by the UK Engineering and Physical Sciences Research Council (EPSRC). Its vision was built on the unique capabilities of gallium nitride (GaN) LEDs which can provide good illumination and high data rate optical communications. New forms of spatial multiplexing were implemented where independent communications channels were provided by individual elements in high-density arrays of GaN based LEDs [199]. The project combined the strengths of researches in the UK in complementary metal-oxide-semiconductors (CMOS), nitrides, organic semiconductors and communications systems engineering to provide a new wireless communication technology that transfers data through visible light instead of RF (Wi-Fi) [200]. Multi-Gb/s VLC was shown as a possibility for the first time by using GaN LEDs and novel multiplexing schemes were explored [200].

Subsequently, color-tunable smart displays were developed using CMOS electronics combined with GaN based micro-LEDs. The modulation bandwidth of the LED pixels of these color-tunable displays is 100 MHz which provides a wavelength-agile source simultaneously for high-speed VLC [199].

2. Visible light based interoperability and networking (VisIoN) project

The VisIoN project started in September 2017 and will conclude in September 2021 and is led by Centrale Marseille in

France. The project focuses on synergizing optical communications technologies with wireless communication technologies. It is built on three key technical work packages (WPs) with a proof-of-concept demonstration for each WP. The first WP focuses on smart cities, offices and homes where everything in the home or city is connected wirelessly. As such, the communication network must provide high data rates, high reliability, high availability and low latency. VLC is proposed in this project instead of RF communication systems as the project expects VLC to provide improvements in terms of the number of devices that can be supported and the achievable capacity. Also, VLC is expected to offer localization and diverse control with motion detection for these devices. The second WP has a focus on smart transportation, with vehicle to vehicle (V2V) and vehicle to infrastructure (V2I) communication through VLC. The last WP considers manufacturing and medical uses of VLC. In manufacturing, VLC can provide unique safety and security features. In medical applications, VLC is expected to offer interference immunity compared to RF. In addition, it is expected that the VLC systems studied will offer data communication, localization and sensing together with high bandwidth, high reliability, robustness and low latency [201].

3. Super Receivers for Visible Light Communications project

The project was established in August 2017 and will conclude July 2020, led by University of Oxford. It focuses on creating a more sensitive new super receiver. This super receiver is designed to overcome sensitivity and data rate limitations and will offer easy alignment. It can be easily integrated onto any device due to its characteristics (thin and flexible). The project plan is to test the receiver in free space visible light communications links to quantify its performance. The project estimates that the new super receiver will lead to a hundred times improvement in the performance compared to the existing receiver by offering higher data transmission rates, longer transmission distance and lower noise [202].

4. Multifunctional Polymer Light-Emitting Diodes with Visible Light Communications (MARVEL) project

The project started December 2016 and is due to conclude November 2019. The project will introduce new Polymer Light-Emitting Diode designs with new approaches that can maximize the light efficiency [203].

5. Terabit Bidirectional Multi-user Optical Wireless System (TOWS) for 6G LiFi project

The project will take place between January 2019 and December 2023, and is led by the University of Leeds working with the Universities of Cambridge and Edinburgh. The project will develop and experimentally demonstrate multiusers Tbps optical wireless systems. The developed systems will provide capacities at least two orders of magnitude higher than the existing and planned 5G optical and radio wireless systems. In addition, a roadmap will be developed to achieve capacities up to four orders of magnitude higher than those of existing 5G wireless systems [204].

6. Designing Future Optical Wireless Communication Networks (DETERMINE) project

The project started in June 2013 and concluded in May 2017, and was led by Linkopings University in Sweden. The project focused on developing concepts, algorithms, models and architectures that exploit the optical spectrum and had a focus on EU exchange of knowledge between the partners of the consortium in the physical layer and the other OW layers [205].

7. Wireless Optical/Radio TErabit Communications (WORTECS) project

The project, with a time window September 2017 to August 2020, led by Orange, France will develop ultra-high data rate wireless systems by combining high radio frequency communications with optical wireless communications in the regions of infrared and visible optical spectrum by utilizing heterogeneous networking concepts. Two Proof-of-Concept demonstrators will be delivered in the project: an ultra-high density LiFi/Radio network with multi-Gbps data rates, and an ultra-high data rate system capable of delivering Tbps networking [206].

B. VLC's products

Light Fidelity (Li-Fi) helped propel OW exploitation bringing it closer to end users and standards. pureLiFi has a commercial product, LiFi-XC, that can transfer data using indoor lights. The system is full duplex and bi-directional. It uses visible light for downlink communication and infrared for the uplink. It can offer up to 43 Mbps with fully networked Li-Fi with mobility allowing users to move around the environment. The pureLiFi device can be separated from, or integrated into the computer [207]. In addition other products include those from Oledcomm which provides a receiver, LiFiNET® Dongle. This receiver can connect the computer to the Internet and can offer a data rate around 10 Mbps [208]. Signify, a new Philips lighting company, offers a Li-Fi system with excellent light quality and 30 Mbps data rate for two way communications [10]. VLNComm provides an advanced Li-Fi adapter called LUMISTICK LIFI ADAPTER. This adapter connects devices to the Li-Fi network offering 108 Mbps for downlink communication and 53 Mbps for uplink communication [209].

IX. OPEN RESEARCH ISSUES

This section summarizes new areas of research that deserve further attention in IRC and VLC systems.

A. Open research issues in IRC systems

Despite the notable solutions introduced for IRC systems to provide mobility, mitigate ISI and achieve high data rates, low complexity and easily implementable techniques are still required. Areas of further investigation in IRC systems are listed below.

- Mobility can degrade the performance of CDS and LSMS systems. However, using certain techniques, such as beam delay adaptation, power adaptation, beam angle adaptation and imaging receivers, can enhance the system performance. However, experimental verification of these techniques has not been carried out yet. Real time adaptation in these systems may lead to very complex solutions. Therefore, investigation is needed to assess these techniques in realistic IRC systems to provide mobility and to achieve multi-gigabit per second data rates. In addition, further investigations are needed to find new approaches to support mobility with minimum complexity.

- Power, angle and beam delay adaptation have been evaluated for a single user scenario. An open area of research is the use of these techniques in multiuser scenario. Multiuser scenarios are necessary for the wide use and adoption of these OW systems.
- In multi-user OW systems, the optimum allocation of resource such as wavelengths, time slots, beams, imaging receiver pixels, and ADR facets is an open research area.
- In the case of a multiuser scenario, scheduling can also be used with the techniques above to maximise throughput, reduce interference, reduce power consumption, increase reliability, reduce delay or optimize with respect to other metrics.
- Efficient use of the optical transmitted power is key in IRC systems due to the power transmitted in IRC systems being restricted by eye safety. Increasing the transmit power can however lead to improved SNR if interference is properly managed. High SNRs can allow simple receiver structures to be used reducing the system complexity. Therefore, effort is required to identify methods that can enable the use of high transmit power in a safe way, such as holographic techniques.
- The optimization of an imaging receiver can lead to low complexity and easier implementation. To the best of our knowledge, there are no high-speed imaging receivers to date that have been specially designed for mobile IRC systems. At high data rates most of the components will probably be adopted from the optical fibre domain, which is not ideal for IRC systems; thus, more research is needed in this area to design and fabricate new devices for IRC systems.
- Applying new techniques, such as OMIMO, in combination with advanced modulation (Optical OFDM and other) may further enhance the data rates achieved in IRC systems.

### B. Open research issues in VLC systems

The main current research focuses on modulation techniques, mobility, uplink connectivity, dimming control and illumination, increasing the achievable data rate, improving the modulation bandwidth of transmitters and mitigating ISI.

- Uplink channel provision is an important issue. The proposed solutions, such as the use of IRC, corner cube reflectors or RF systems, unfortunately do not fit well with the VLC system goal to provide high data rate in the uplink and down link as these techniques may conflict with providing mobility and minimizing the size of the VLC devices (portable device). Therefore, new schemes to provide an uplink for VLC are required. Intermediate stations (relays) have been studied previously for use in a variety of networks to increase the coverage area and the capacity. Relay stations may be used for VLC systems to provide an up link connection for users and enable communication between users.
- Further studies are necessary to answer questions relating to the optimum dimming control and optimum modulation for VLC that provide a good balance between communication and illumination.
- Providing mobility for VLC systems is vital for indoor users. The ADR and imaging receiver evaluated in IRC systems can be used to offer mobility for VLC systems and to mitigate the ISI problem at the same time. Our previous work in this area has shown that significant enhancements can be achieved when using an imaging receiver [24].
- LDs can be used for illumination and communication in VLC systems instead of LEDs in order to enable higher data rates. Initial results have shown that data rates of tens of gigabits per second per user and more can be achieved [10], [24], [25]. Further work is needed in this area to optimize the transmitter and receiver devices, their properties and architecture.
- Due to the energy efficiency, scalability and flexibility of VLC systems, they can be used in next generation data centres replacing hundreds of metres of cables or optical fibre. Investigations in this area are required to examine the benefits of applying VLC systems in data centres.
- The main frontier in VLC and OWC now is networking. This includes networking and MAC techniques applied in the optical medium to optimize beams, receiver detectors fields of view and orientations, and to determine the optimal allocation of resources that include wavelengths, space and time resources. The optimum front haul and backhaul network architectures are open questions that deserve substantial attention to deliver and collect data from optical luminaires or from IRC sources and detectors.
- The use of machine learning and artificial intelligence in VLC and OWC is important and timely to carry out the optimum resource allocation described above, to enable the network to self-configure and self-adapt taking into account interference, capacity, power consumption, throughput and reliability and availability into account.

## X. CONCLUSIONS

In this paper, we have presented an overview of OW systems that emphasized their deployment to provide indoor communication. We illustrated how OW technologies can offer a complementary technology for future wireless communications in indoor applications, co-existing potentially with RF systems.

Two distinct categories of OW systems (IRC and VLC) systems were discussed. Link design, modulation techniques, transmitter/receiver structures, possible challenges, the main applications and open research issues for each category have been identified.

IRC systems can provide extremely high data rates and can support mobility. These are however still significantly challenging due to the indoor OW channel nature. However, suitable techniques such as the use of the space dimension through the introduction of suitable beams, beam angle, beam power adaptation and beam delay adaptation, together with the use of imaging receivers can be considered to provide mobility and enhance the performance of the indoor IRC systems.

VLC systems are among the promising solutions to the bandwidth limitation problem faced by RF systems. Significant research effort is being directed towards the development of VLC systems due to their numerous advantages over radio systems. Networking in VLC systems is a key outstanding challenge.

## XI. ACKNOWLEDGMENTS

We would like to acknowledge funding from the Engineering and Physical Sciences Research Council (EPSRC) for the INTERNET (EP/H040536/1) and STAR (EP/K016873/1) projects. All data are provided in full in the results section of this paper. The authors extend their appreciation to the International Scientific Partnership Program ISPP at King Saud University, Kingdom of Saudi Arabia for funding this research work through ISPP#0093.